\magnification=\magstep1
\tolerance500
\rightline{Fermilab-Pub-97/227-T}
\rightline{ TAUP 2409-97}
\rightline{Revised 7 April 1998}
\vskip 3 true cm
\centerline {\bf Representation of Quantum Mechanical Resonances}
\centerline{\bf in the}
\centerline{\bf Lax-Phillips Hilbert Space}
\vskip 1 true cm
\centerline{ Y. Strauss$^\dagger$ and L.P. Horwitz$^{\ast,\dagger,
\ddagger}$}
\centerline{E. Eisenberg$^\ddagger$}
\centerline{$^\ast$Fermi National Accelerator Laboratory}
\centerline{Box 500 Batavia, IL 60510}
\centerline{$^\dagger$School of Physics, Tel Aviv University}
\centerline{Raymond and Beverly Sackler Faculty of Exact Sciences}
\centerline{Ramat Aviv, 69978 Israel}
\centerline{$^\ddagger$Department of Physics}
\centerline{Bar Ilan University}
\centerline{ Ramat Gan, 59500 Israel}
\vskip 2 true cm
\noindent
{\it Abstract:\/}  We discuss the quantum Lax-Phillips theory of
 scattering and unstable systems.  In this framework, the decay of an
unstable system is described by a semigroup.  The spectrum of the
generator of the semigroup corresponds to the singularities of the
Lax-Phillips $S$-matrix. In the case of discrete (complex) spectrum of
the generator of the semigroup,
associated with resonances, the decay
law is exactly exponential.  The states corresponding to these
resonances (eigenfunctions of the generator of the semigroup)
lie in the Lax-Phillips Hilbert space, and therefore all physical
properties of the resonant states can be computed.
  We show that the
Lax-Phillips $S$-matrix is unitarily related to the $S$-matrix of standard
scattering theory by a unitary transformation parametrized by the spectral
variable $\sigma$ of the Lax-Phillips theory. Analytic continuation in
$\sigma$ has some of the properties of
 a method developed some time
ago for application to dilation analytic potentials.
  We work out an illustrative example using
 a Lee-Friedrichs model for the underlying dynamical system.
 \vfill
\eject
\noindent
{\bf 1. Introduction}
\smallskip
\par There has been considerable effort in recent years in the development
of the theoretical framework of Lax and Phillips scattering theory$^1$
for the description of quantum mechanical systems$^{2,3,4}$. This work
is motivated by the requirement that the decay law of a decaying system
should be exactly exponential if the simple idea that a set of independent
unstable systems consists of a population for which each element has
a probability, say $\Gamma$, to decay, per unit time. The resulting
exponential law ($\propto e^{-\Gamma t}$) corresponds to an exact semigroup
 evolution of the
state in the underlying Hilbert space, defined as a family of bounded
operators on that space satisfying
$$ Z(t_1) Z(t_2)= Z(t_1+t_2), \eqno(1.1)$$
where $t_1,\, t_2 \geq 0$, and $Z(t)$ may have no inverse. If the
decay of an unstable system is to be associated with an irreversible
process, then its evolution necessarily has the property $(1.1)$.$^5$
The standard model of Wigner and Weisskopf$^6$,
 based on the computation of the survival
amplitude $A(t)$ as the scalar product
$$ A(t) = (\psi, e^{-iHt} \psi) \eqno(1.2)$$
 where $\psi$ is the initial state of the unstable system and $H$ is
the Hamiltonian for the full evolution, results in a good
approximation to an exponential decay law for values of $t$
sufficiently large
(Wigner and Weisskopf$^6$ calculated an atomic linewidth in this
way) but cannot result in a semigroup$^7$.
\footnote{*}{This formula is generally applied to the transitions
 induced by interacting
fields on states in the Hilbert space of a quantum field theory as well.}
When applied to a two-channel system, such as the decay of the $K^0$
meson, one easily sees that the poles of the resolvent for the Wigner-Weisskopf
evolution of the two channel systems result in non-orthogonal residues
that generate interference terms, which destroy the semigroup property,
 to accumulate in the calculation of predictions for
regeneration experiments$^8$.
The Yang-Wu$^9$ parametrization of the $K^0$ decay processes, based on a
Gamow$^{10}$ type evolution generated by an effective $2x2$ non-Hermitian
matrix Hamiltonian, on the other hand, results in an evolution that is an
exact semigroup. It should be possible, with sufficiently careful
experiments, to observe the difference in the two types of
predictions. It appears that the phenomenological parametrization
of refs. 9, which results in semigroup evolution, is indeed
consistent to a high degree of accuracy with the experimental
results on $K$-meson decay$^{11}$.
\par  The Wigner-Weisskopf model results in non-semigroup
evolution independently of the dynamics of the system. Reversible
transitions of a quantum mechanical system, such as adiabatic
precession of a magnetic moment or tunneling through a potential barrier$^{12}$,
which are not radiative, could be expected to be well-described by the
Wigner-Weisskopf formula.
\par In order to achieve exact exponential decay, methods of
analytic extension of the Wigner-Weisskopf model to a generalized
 space have been studied$^{13}$. The generalized states, occurring in
the large sector of a Gel'fand triple, are constructed by defining a
bilinear form, and analytically
continuing a parameter (energy eigenvalue) in one of the vectors to
achieve an exact complex eigenvalue.  Although it is possible to achieve
an exact exponential decay in this way, the resulting (Banach space) vector
has no
properties other than to describe this decay law; one cannot
compute other properties of the system in this ``state''.  Identifying
some representation of the resonant state, it would be of
interest, in some applications, to compute, for example, its
localization properties, its momentum distribution, or its mean
spin.
\par  The quantum Lax-Phillips theory$^{2,3}$, constructed by
embedding the quantum theory into the original Lax-Phillips scattering
theory$^1$ (originally developed for hyperbolic
systems, such as acoustic or electromagnetic waves),  describes the resonance
 as a
state in a Hilbert space, and therefore it is possible, in principle,
to calculate all measurable properties of the system in this state.
Moreover, the quantum Lax-Phillips theory provides a framework for
understanding the decay of an unstable system as an irreversible
process.  It appears, in fact, that this framework is categorical for
the description of irreversible processes.
\par It is clearly desirable to construct a theory which
admits the exact semigroup property, but has sufficient structure to
describe non-semigroup behavior as well, according to the dynamical
properties of the system. The quantum Lax-Phillips theory contains the
latter possibility as well, but in this work, we shall restrict ourselves
to a study of the semigroup property, associated with irreversible
processes.
\par The scattering theory of Lax and Phillips assumes the existence
of a Hilbert space $\overline{\cal H}$ of physical states in which there are two
 distinguished orthogonal subspaces ${\cal D}_+$ and ${\cal D}_-$ with the
 properties
$$\eqalign{ U(\tau)\,{\cal  D}_+ &\subset \,{\cal D}_+  \qquad \tau > 0 \cr
U(\tau) \,{\cal D}_- &\subset \,{\cal D}_- \qquad \tau < 0\cr
\bigcap_\tau \,U(\tau)\,{\cal D}_\pm &= \{0\} \cr
\overline{\bigcup_\tau \,U(\tau)\,{\cal D}_\pm }
&= \overline {\cal H}, \cr} \eqno(1.3)$$
i.e., the subspaces ${\cal D}_\pm$ are stable under the action of the full
 unitary dynamical evolution $U(\tau)$, a function of the physical
 laboratory time, for positive and negatives
times $\tau$ respectively; over all $\tau$, the evolution operator
generates a dense
 set in $\overline{\cal H}$ from either ${\cal D}_+$ or ${\cal D}_-$.
 We shall call ${\cal D}_+$ the {\it outgoing subspace} and
${\cal D}_-$ the {\it incoming subspace} with respect to the group $U(\tau)$.
\par A theorem of Sinai$^{14}$ then assures that $\overline {\cal H}$ can
be represented as a family of Hilbert spaces obtained by foliating
$\overline {\cal H}$ along the real line, which we shall call
$\{t\}$, in the form of a direct integral
$$\overline{\cal H} = \int_\oplus {\cal H}_t,   \eqno(1.4)$$
where the set of auxiliary Hilbert spaces ${\cal H}_t$ are all
isomorphic.  Representing these spaces in terms of square-integrable
functions, we define the norm in the direct integral space (we use
Lesbesgue measure) as
$$ \Vert f \Vert^2 = \int_{-\infty}^\infty \,dt \Vert f_t \Vert^2_H,
\eqno(1.5)$$
where $f \in \overline{H}$ represents
$\overline {\cal H} $ in terms of  the $L^2$ function space
$L^2(-\infty,\infty, H)$,
and $f_t \in H$, the $L^2$ function space
representing ${\cal H}_t$ for any $t$. The Sinai theorem
 furthermore asserts that
there are representations for which the action of the full
evolution group $U(\tau)$ on $L^2(-\infty,\infty, H)$ is translation
by $\tau$ units. Given $D_\pm$ (the $L^2$ spaces representing ${\cal D}_\pm$),
 there is such a representation, called the
{\it incoming representation}$^1$, for which functions in $D_-$ have support
in $L^2(-\infty,0, H)$, and another called the {\it outgoing
representation}, for which functions in $D_+$ have support in $L^2(0,
 \infty,H)$.
\par  Lax and Phillips$^1$ show that there are unitary operators
$W_\pm$,
 called
 wave
operators, which map elements in $\overline{\cal H}$,
respectively, to these representations.  They define an $S$-matrix,
$$ S= W_+W_-^{-1} \eqno(1.6)$$
which connects these representations; it is unitary, commutes with
translations, and maps $L^2(-\infty,0)$ into itself.  The
singularities of this $S$-matrix, in what we shall define as the {\it
spectral representation}, correspond to the spectrum of the generator
of the exact semigroup characterizing the evolution of the unstable
system.
\par With the assumptions stated above on the properties of the
subspaces ${\cal D}_+$ and ${\cal D}_-$, Lax and Phillips$^1$ prove that the
 family
of operators
$$ Z(\tau) \equiv P_+ U(\tau) P_- \qquad (\tau \geq 0), \eqno(1.7)$$
where $P_\pm$ are projections into the orthogonal complements of
${\cal D}_\pm$, respectively, is a contractive, continuous, semigroup.  This
 operator
 annihilates vectors in ${\cal D}_\pm$ and carries the space
$$ {\cal K} = \overline {\cal H} \ominus {\cal D}_+ \ominus
 {\cal D}_- \eqno(1.8)$$
into itself, with norm tending to zero for every element in
${\cal K}$.
\par The existence of a semigroup law for
transitions in the framework of the usual quantum mechanical Hilbert space
has been shown to be unattainable$^7$.  However, Flesia and Piron$^2$
 found that the {\it direct integral} of quantum mechanical Hilbert spaces
 can provide a framework for the Lax-Phillips construction,
resulting in a structure directly analogous to the
foliation $(1.4)$.  In this
construction, it was found$^4$ that for the representation in which
the free evolution is represented by translation on the foliation
parameter in Eq. $(1.5)$ (and for which it is assumed that $D_\pm$ have
definite support properties), the full
evolution of the system must be an integral kernel in order to achieve
the connection between the Lax-Phillips
$S$-matrix and the semigroup. In this work we show that the evolution
operator for the  physical model
for the system may be pointwise, in a representation
which we shall call the {\it model representation}, but in another
representation, corresponding to a different foliation, the necessary
conditions for the construction of a non-trivial Lax-Phillips theory
can be naturally realized.
 The natural association of
the time parameter in the model representation  with the foliation asserted
 by the theorem of
Sinai$^{14}$, as we shall show, does not
necessarily correspond to the proper embedding of the quantum theory
into the Lax-Phillips framework.
\par If we identify elements in the space ${\overline {\cal H}}$ with
{\it physical states}, and identify the subspace ${\cal K}$ with
 the unstable system, we see
that the quantum Lax Phillips theory provides a framework for the description
of an unstable system which decays according to a semigroup law. We
remark that, taking a vector $\psi_0$ in ${\cal K}$, and evolving it under
the action of $U(\tau)$, the projection back into the original state
 is\footnote{\ddag}{It follows from $(1.7)$ and the stability of
${\cal D}_\pm$
 that $Z(\tau) = P_{\cal K}U(\tau)P_{\cal K}$ as well.}
$$ \eqalign{A(\tau) &= (\psi_0, U(\tau) \psi_0) \cr
                    &= (\psi_0, P_{\cal K} U(\tau)P_{\cal K} \psi_0) \cr
                    &= (\psi_0, Z(\tau) \psi_0), \cr }\eqno(1.9)$$
so that the survival amplitude of the Lax-Phillips theory, analogous
to that of the Wigner-Weisskopf formula $(1.2)$, has the
exact exponential behavior. The difference between this result and the
corresponding expression $(1.2)$ for the Wigner-Weisskopf theory can
be accounted for by the fact that there are translation
representations for $U(\tau)$, and that the definition of the subspace
${\cal K}$ is related to the support properties along the foliation
axis on which these translations are induced.$^3$
\par  Functions in the  space ${\overline H}$, representing the
elements of  ${\overline{\cal H}}$,  depend on the variable $t$ as
well as the variables of the auxiliary space $H$. The measure space
 of this Hilbert space of states is one dimension larger than that of a
quantum theory represented in the auxiliary space alone.  Identifying
this additional variable with an {\it observable} (in the sense of a
 quantum mechanical observable) time, we may understand this
representation of a state as a {\it virtual history}. The collection
of such histories forms a quantum ensemble; the absolute square of the
wave function corresponds to the probability that the system
would be found, as a result of measurement, at time $t$
 in a particular configuration in the auxiliary space (in the
state described by this wave function), i.e., an element of
one of the virtual histories.   For example, the expectation value of
the
 position variable  $x$ at a given $t$ is, in the standard
interpretation
 of the auxiliary space as a space of quantum states,
$$ \langle x \rangle_t = {(\psi_t, x \psi_t) \over \Vert \psi_t
 \Vert^2}. \eqno(1.10)$$
The full expectation value in the physical Lax-Phillips state,
according to $(1.5)$,
 is then$^4$
$$ \int dt \, (\psi_t, x \psi_t) = \int dt\, \Vert \psi_t \Vert^2
\langle x
 \rangle_t, \eqno(1.11)$$
so we see that $\Vert \psi_t\Vert^2$ corresponds to the probability to
find a signal which indicates the presence of the system at the time
$t$ (in the same way that $x$ is interpreted as a dynamical variable
in the quantum theory).
\par One may ask, in this framework, which results in a precise
semigroup behavior for an unstable system, whether such a theory can
support as well the description of stable systems or a system which
makes a transition following the rule of Wigner and Weisskopf (as, for
example, the adiabatic rotation of an atom with spin in an electromagnetic
field).  It is clear that if $D_\pm$ span the whole space, for example,
 there is no unstable
subspace, and one has a scattering theory without the type of
resonances that can be associated with unstable systems. We
 shall treat this subject
in more detail in a succeeding paper.
\par In the next section, we give a
procedure for the construction of the subspaces $D_\pm$, and for
defining the representations which realize the Lax-Phillips structure.
In this framework, we define the Lax-Phillips $S$-matrix.
In Section 3, we show that this construction results in a Lax-Phillips
theory applicable to models in which the underlying dynamics is
locally defined in time.  We carry out the construction for a
     Flesia-Piron type model.  In Section 4 we study the general form
of the Lax-Phillips $S$-matrix and prove that it is unitarily related to
the standard $S$-matrix of the usual scattering theory in the auxiliary
space.   In Section 5,  we work out the specific example of a time-independent
Lee-Friedrichs spectral model$^{15}$,
 and show that the condition for the resonance eigenfunction is closely
 related to
the resonance pole condition of the Lee-Friedrichs model of the usual quantum
 theory.
 A discussion
and conclusions are given in Section 6.
\bigskip
{\bf 2. The Subspaces ${\cal D}_\pm$, Representations,
and the Lax-Phillips $S$-Matrix}
\smallskip
 \par It follows from the existence of the
one-parameter unitary group $U(\tau)$ which acts on the Hilbert space
$\overline{\cal H}$ that there is an operator $K$ which is the
generator of  dynamical evolution of the physical states in $\overline{\cal
H}$; we assume that there exist {\it wave operators} $\Omega_\pm$ which
intertwine this dynamical operator with an unperturbed dynamical
operator
$K_0$.
 We shall assume that $K_0$ has only absolutely
continuous spectrum in $(-\infty,\, \infty)$.
\par We begin the development of the quantum Lax-Phillips theory with
the construction of these representations. In this way, we shall
construct explicitly the foliations described in Section 1.  The
{\it free spectral representation} of $K_0$ is defined by
$$   _f\langle \sigma \beta \vert K_0 \vert g \rangle = \sigma\,_f\langle
\sigma \beta \vert g \rangle , \eqno(2.1)$$
where $\vert g \rangle$ is an element of $\overline{\cal H}$ and
$\beta$ corresponds to the variables (measure space) of the auxiliary
space associated
to each value of $\sigma$, which, with $\sigma$, comprise a complete
spectral set.  The functions $_f\langle \sigma \beta
\vert g \rangle$ may be thought of as a set of functions of the
variables $\beta$ indexed on the variable
 $\sigma$ in a continuous sequence of auxiliary  Hilbert spaces
isomorphic to $H$ .

\par We now proceed to define the incoming and outgoing subspaces
$\cal D_\pm$.  To do this, we define the Fourier transform from
representations according to the spectrum $\sigma$ to the foliation
variable $t$ of $(1.5)$, i.e.,
$$ _f\langle t \beta \vert g \rangle =  \int e^{i\sigma t}\  _f\langle
\sigma \beta \vert g \rangle d\sigma . \eqno(2.2)$$
 Clearly, $K_0$  acts as
the generator of translations in this representation. We shall say
that the set of functions $ _f\langle t \beta \vert g \rangle $ are in
the {\it free translation representation}.
\par  Let us consider
the sets of functions with support in $L^2(0, \infty)$ and in
$L^2(-\infty,0)$,and call these subspaces $ D_0^\pm$.  The
Fourier transform back to the free spectral representation provides
the two sets of Hardy class functions
$$ _f\langle \sigma \beta \vert g_0^\pm \rangle =  \int e^{-i\sigma
t}\
 _f\langle t \beta \vert g_0^\pm \rangle dt  \in H_\pm , \eqno(2.3)$$
for $g_0^\pm \in D_0^\pm$.
\par We may now define the subspaces ${\cal D}_\pm$ in the Hilbert space of
states ${\overline{\cal H}}$.  To do this we first map these Hardy
class functions in ${\overline H}$ to ${\overline{\cal H}}$, i.e., we
define the subspaces ${\cal D}_0^{\pm}$ by
$$\int \sum_\beta \vert\sigma \beta \rangle_f \  _f\langle \sigma
\beta \vert g_0^\pm \rangle d\sigma \in {\cal D}_0^\pm.  \eqno(2.4)$$
\par   We shall assume that there are wave operators which intertwine $K_0$
with the full evolution $K$, i.e., that the limits
$$ \lim_{\tau \to \pm \infty} e^{iK\tau} e^{-iK_0 \tau} = \,
\Omega_\pm \eqno(2.5)$$
exist on a dense set in ${\overline{\cal H}}$.\footnote{\S}{We
emphasize that the
 operator $K$
generates evolution of the entire virtual history, i.e., of elements
in $\overline {\cal H}$, and that these wave operators are defined in this
larger space. These operators are {\it not},in general,
 the usual wave (intertwining) operators for the perturbed and
unperturbed Hamiltonians that act in the auxiliary space.  The
conditions for their existence are, however,
 closely related to those of the usual wave operators. For the
existence of the
 limit, it is sufficient that for $\tau \rightarrow \pm \infty, \,
 \Vert V e^{-iK_0\tau} \phi \Vert \rightarrow 0$ for a dense set
in $\overline {\cal H}$.  The free evolution may induce a motion of
the wave packet in the auxiliary space out of the range of the potential (in
the variables of the auxiliary space in the model representation),
 as for the usual scattering theory, so
that it is possible to construct examples for which the wave operator exists
if the potential falls off sufficiently rapidly.}
\par  The construction of ${\cal D}_\pm $ is then completed with the help
of the wave operators.  We define these subspaces by
$$\eqalign{ {\cal D}_+ &= \Omega_+ {\cal D}_0^+  \cr
{\cal D}_- &= \Omega_- {\cal D}_0^-  .\cr} \eqno(2.6)$$
We remark that these subspaces are not produced by the same unitary
map. This procedure is necessary to realize the Lax-Phillips structure
non-trivially; if a single unitary map were used, then there would
exist a transformation into the space of functions on $L^2(-\infty,
\infty, H)$ which has the property that all functions with support on
the positive half-line represent elements of ${\cal D}_+$, and all
functions with support on the negative half-line represent elements of
${\cal D}_-$ in the same representation; the resulting Lax-Phillips
$S$-matrix would then be trivial.
 The requirement that ${\cal D}_+$
and ${\cal D}_-$ be orthogonal is not an immediate consequence of our
construction; as we shall see, this result is associated with the
analyticity of the operator which corresponds to the Lax-Phillips
 $S$-matrix.
\par In the following, we construct the Lax-Phillips $S$-matrix and
the Lax-Phillips wave operators.
\par The wave operators defined by $(2.5)$ intertwine $K$ and $K_0$, i.e.,
$$ K \Omega_\pm  = \Omega_\pm K_0;        \eqno(2.7)$$
we may therefore construct the outgoing (incoming) spectral representations
from the free spectral representation.  Since
$$ \eqalign{K\Omega_\pm \vert \sigma \beta \rangle_f &= \Omega_\pm K_0 \vert
\sigma \beta \rangle_f \cr
&=\sigma \Omega _\pm \vert \sigma \beta \rangle_f,\cr} \eqno(2.8)$$
we may identify
$$ \vert \sigma \beta \rangle_{out \atop in} = \Omega_\pm \vert \sigma
\beta \rangle_f . \eqno(2.9)$$
The Lax-Phillips $S$-matrix is defined as the operator, on ${\overline H}$,
which carries the incoming to outgoing translation representations of the
evolution operator $K$. Suppose $g$ is an element of ${\overline {\cal H}}$;
its incoming spectral representation, according to $(2.9)$, is
$$ {_{in}\langle} \sigma \beta \vert g) = {_f\langle}\sigma \beta \vert
\Omega_-^{-1} g).  \eqno(2.10)$$
Let us now act on this function with the Lax-Phillips $S$-matrix in the
free spectral representation, and require the result to be the {\it outgoing}
representer of $g$:
$$ \eqalign{{_{out}\langle} \sigma \beta \vert g)
&= {_f\langle} \sigma \beta \vert \Omega_+^{-1} g) \cr
&=\, \int d\sigma'\, \sum_{\beta'}{_f\langle} \sigma \beta \vert
 {\bf S}\vert \sigma'\beta' \rangle_f \,\,
{_f\langle}\sigma' \beta' \vert \Omega_-^{-1} g) \cr} \eqno(2.11)$$
where ${\bf S}$ is the Lax-Phillips $S$-operator
 (defined on ${\overline{\cal H}}$).
Transforming the kernel to the free translation representation
with the help of
$(2.2)$, i.e.,
$$ {_f\langle} t \beta\vert {\bf S} \vert t' \beta' \rangle_f =
{1 \over (2\pi)^2}
\int d\sigma d\sigma' \, e^{i\sigma t} e^{-i\sigma't'}
{_f\langle} \sigma \beta \vert
 {\bf S}\vert \sigma'\beta' \rangle_f , \eqno(2.12)$$
we see that the relation  $(2.11)$ becomes, after using Fourier
transform in a similar way to
transform the {\it in} and {\it out } spectral representations to
the corresponding {\it in} and {\it out} translation representations,
$$\eqalign{ {_{out}\langle} t\beta \vert g) = {_f\langle} t\beta
 \vert \Omega_+^{-1} g) &=
\int dt'\, \sum_{\beta'} {_f\langle} t \beta\vert {\bf S} \vert t' \beta'
\rangle_f
\,{_f\langle} t' \beta' \vert \Omega_-^{-1} g) \cr
&=  \int dt'\, \sum_{\beta'} {_f\langle} t \beta\vert {\bf S} \vert t' \beta'
 \rangle_f {_{in}\langle} t'\beta' \vert g). \cr}  \eqno(2.13)$$
Hence the Lax-Phillips $S$-matrix is given by
$$ S= \{ {_f\langle} t \beta\vert {\bf S} \vert t' \beta' \rangle_f
\},
 \eqno(2.14)$$
 in free translation representation. It follows from the intertwining
property $(2.7)$ that
$$ {_f\langle} \sigma \beta \vert{\bf S}\vert \sigma' \beta' \rangle_f =
\delta(\sigma - \sigma') S^{\beta \beta'}(\sigma), \eqno(2.15)$$
 \par This result can be expressed in terms of operators on ${\overline{\cal
H}}$.
  Let
$$ w_-^{-1} = \{ {_f\langle} t\beta \vert \Omega_-^{-1} \}   \eqno(2.16)$$
be a map from ${\overline{\cal H}}$ to ${\overline H}$ in the incoming
 translation
representation, and, similarly,
$$ w_+^{-1} = \{ {_f\langle} t\beta \vert \Omega_+^{-1} \} \eqno(2.17)$$
a map from ${\overline{\cal H}}$ to ${\overline H}$ in the outgoing translation
representation. It then follows from $(2.13)$ that
$$ S= w_+^{-1} w
_- , \eqno(2.18)$$
as a kernel on the free translation representation.
 This kernel is
understood to operate on the representer of a vector $g$ in the incoming
representation and map it to the representer in the outgoing representation.

\par We now discuss a class of pointwise physical models, and return
in Section 4 to the construction of the Lax-Phillips $S$-matrix for this
class of models.

\bigskip
\noindent
{\bf 3. Pointwise Physical Models}
\smallskip
\par It has been shown by Piron$^5$
that if\footnote{\dag}{The  symbol $-i\partial_t$
stands, in this context,  for the operator on $\overline{\cal H}$
 which acts on the
family $\{{\cal H}_t\}$ as a partial derivative in the foliation
parameter.} $K$, $-i\partial_t$,
 and $K+i\partial_t$ have a common dense
domain on which they are essentially self-adjoint, then there exists
an operator H, defined
as the self-adjoint extension of $K+i\partial_t$, which  is a decomposable
operator on $\overline H$, that is, $({\rm H}\psi)_t = {\rm
H}_t\psi_t$.
 We therefore have, on this common domain,
$$ K= -i\partial_t + {\rm H} , \eqno(3.1)$$
corresponding to an evolution which acts pointwise in $t$.
We shall identify the representation in which this analysis is carried
out with what we have called the {\it model representation}.
\par In this section,
 we show that physical models of this type, for which the evolution
is defined
pointwise in time (in the model representation), which provide a
straightforward way of lifting problems in the framework of the usual
quantum theory to the Lax-Phillips structure,  satisfy the
requirements imposed by Eisenberg and Horwitz$^4$ on the structure
 of a nontrivial Lax-Phillips theory,
 i.e., that the evolution be
represented by a nontrivial kernel in the free translation representation.
\par Consider  a class of models for nonrelativistic quantum theory
characterized
by the standard Heisenberg equations\footnote{*}{Context should avoid confusion
between the symbol $\rm H$ for the Hamiltonian and the designation of the
auxiliary Hilbert space $H$.}
$$ {d{\bf x} \over dt} = i[{\rm H}, {\bf x}] \qquad {d{\bf p} \over dt} =
 i[{\rm H}, {\bf p}] \eqno(3.2)$$
in terms of operators defined on a Hilbert space $H$,
where
$$ {\rm H} = {\rm H}_0 + V. \eqno(3.3)$$
In case there is an explicit time dependence in $V= V(t)$, for example, in a
model
in which the interaction that induces instability is turned on at some
finite laboratory time, it is often convenient to
formally adjoin two new dynamical variables (as done, for example, by Piron$^5$
and Howland$^{16}$), $T_m$ and $E$, along with an evolution
 parameter $\tau$
 to replace the role of the parameter $t$ in $(3.2)$ ($T_m$  denotes the
time operator in the space in which we construct
the dynamical model of the system; such a time operator exists because the
spectrum
of $E$ is taken to be $(-\infty, \infty)$).  The evolution operator may then be
 considered
 ``time'' ($\tau$)-independent, i.e., we define, as operators on a larger space
$\overline{\cal H}$ (and thus identify {\rm H} with the decomposable
 operator in $(3.1$)
$$ K= E+{\rm H} = K_0 +V, \eqno(3.4)$$
where
$$K_0 = E+{\rm H}_0, \eqno(3.5)$$
and
$$[T_m, E] = i. \eqno(3.6)$$
Then, Eqs.$(3.2)$ become
$$\eqalign{{d{\bf x} \over d\tau} &= i[K,{\bf x}] = i[{\rm H},{\bf x}] \cr
{d{\bf p} \over d\tau} &= i[K,{\bf p}] = i[{\rm H},{\bf p}] \cr}\eqno(3.7)$$
and
$$\eqalign{{dE \over d\tau }&= i[K,E] = i[{\rm H}, E] \cr
{dT_m \over d\tau} &= i[K, T_m] = i[E, T_m] = 1.
\cr} \eqno(3.8) $$
The first of $(3.8)$ implies, since ${\rm H}_0$ is independent of $t$,
 that$^{16}$
$$ {dE \over d\tau} = -{\partial V \over \partial t}, \eqno(3.9)$$
and the last of $(3.8)$ puts $t$ and $\tau$ into correspondence, i.e.,
the expectation value of $t$ goes with $\tau$.  The evolution of the
system is, however, generated by the operator
$$U(\tau) = e^{-iK\tau}, \eqno(3.10)$$
corresponding to the Lax-Phillips evolution assumed in $(1.3)$. The extension
 we have constructed (by
the inclusion of the operators $T_m$ and $E$) enables us to embed the
 time-dependent
non-relativistic Heisenberg equations into the Lax-Phillips theory, in a way
equivalent to the Flesia-Piron direct integral.  The conditions that they
 impose, that
$E$ and $K$ have a common dense domain, results, by means of the Trotter
 formula${}$, in the
conclusion that $\rm H$ acts pointwise in the spectral decompositon of
 $T_m$.  This result gives $(3.4)$ a precise meaning.  That $K_0$ shares this
common
 domain follows from the
requirement that $V$ be ``small'' $^{17}$.
\par  We shall label
the spectral representation of the operator $T_m$ by the subscript $m$,
 so that for
$\psi \in {\overline{\cal H}}$,
$$ {_m\langle} t\alpha \vert K_0 \vert \psi) = -i\partial_t \,
 {_m\langle} t\alpha
 \vert
\psi) +\,  {_m\langle} t \alpha \vert {\rm H_0} \vert \psi), \eqno(3.11)$$
where $\{\alpha\}$ corresponds to a complete set in the (auxiliary) Hilbert
space associated to $t$.  We shall assume that ${\rm H}_0$ has no $t$
 dependence,
and that $V$ is {\it diagonal} in $t$, so that
$$ {_m\langle t} \alpha \vert {\rm H}_0 \vert \psi) = \sum_{\alpha'}
 {\rm H}_0 ^{\alpha,
\alpha'}\,{_m\langle t} \alpha' \vert \psi)   \eqno(3.12)$$
and
$${_m\langle} t \alpha \vert V\vert \psi) = \sum_{\alpha'} V^{\alpha,
\alpha'}
(t)
\, {_m\langle} t \alpha' \vert \psi).  \eqno(3.13)$$
We therefore see explicitly that the Hilbert space associated to the action
 of the
operator $\rm H$ may be identified with
the auxiliary space of the Lax-Phillips
theory, and the larger space, representing the action of $T_m$ and
$E$,
 with the
function space $\overline H$ or the abstract space ${\overline{\cal H}}$ of
 the
Lax-Phillips theory, as in the (direct integral) construction of Flesia and
 Piron${}$.
\par The free  spectral
representation discussed in Section 2 is constructed
by requiring that $K_0$, in this
representation, act as multiplication.  As in $(2.1)$, we label this
representation with
subscript $f$, and require, for $\psi \in {\overline{\cal H}}$,
$$ _f\langle \sigma \beta \vert K_0\vert  \psi) = \sigma _f\langle \sigma \beta
 \vert \psi),
\eqno(3.14)$$
where $\{\beta\}$ corresponds to a complete set in the (auxiliary) Hilbert
 space
associated to $\sigma$.  This relation defines the free spectral
 representation.
\par The free translation representation is then given by $(2.2)$,
i.e.,
$$ _f\langle t \beta \vert \psi) = \int_{-\infty}^\infty
 e^{i\sigma t}\,{ _f\langle} \sigma \beta
\vert \psi ) d\sigma. \eqno(3.15)$$
 One obtains,
 from $(3.11)$-$(3.14)$, the relation
$$\eqalign{_m\langle t\alpha \vert K_0 \vert \sigma \beta \rangle_f &=
 \sigma\, { _m\langle} t \alpha
\vert \sigma \beta \rangle_f \cr
&= -i \partial_t \,{_m\langle} t\alpha \vert \sigma \beta \rangle_f +
 \sum_{\alpha '} H_0^{\alpha
\alpha'}\, {_m\langle} t \alpha' \vert \sigma \beta \rangle_f. \cr}\eqno(3.16)$$
Making the transformation
$$ _m\langle t \alpha \vert \sigma \beta \rangle_f =
 e^{i\sigma t}\,{ _m^0\langle} t \alpha \vert
\sigma \beta \rangle_f , \eqno(3.17) $$
the relation $(3.16)$ becomes
$$ i \partial_t\,\, { _m^0\langle} t \alpha \vert \sigma \beta\rangle_f =
\sum_{\alpha'}{\rm  H}_0^{\alpha \alpha'}
\,{_m^0\langle} t \alpha' \vert \sigma \beta \rangle_f, \eqno(3.18)$$
or
$$ _m^0\langle t \alpha \vert \sigma \beta \rangle_f
= \sum_{\alpha'} \bigl( e^{-i{\rm H}_0 t}
 \bigr)^{\alpha \alpha'}
\,{_m^0\langle} 0 \alpha' \vert \sigma \beta\rangle_f . \eqno(3.19)$$
The solution $(3.19)$ of $(3.18)$ is norm-preserving in $H$, and therefore
 $_m^0\langle t \alpha \vert
\sigma \beta\rangle_f$ are not elements of ${\overline H}$ (the integral of
 the modulus squared over
$t$ diverges). This norm-preserving evolution reflects the stability
 of the system under
evolution induced by ${\rm H}_0$.  The factor $e^{i\sigma t}$ in $(3.17)$
 imbeds physical
states into $\overline H$.  To see this, consider the norm of $_m\langle t
 \alpha \vert \psi)$,
$$ \eqalign{ \int dt \sum_\alpha \vert _m\langle t \alpha  \vert \psi)
 \vert^2 &=
\int d\sigma\,d\sigma'\,dt \sum_{\alpha\beta\beta'} e^{-i(\sigma -
 \sigma')t}\,{ _m^0\langle}
t \alpha \vert \sigma \beta \rangle_f^* \cdot \cr
&\qquad \cdot { _m^0\langle} t \alpha
\vert \sigma' \beta' \rangle_f\,\,
_f\langle \sigma \beta \vert \psi)^*\, _f\langle \sigma' \beta' \vert \psi) \cr
&= \int dt\, d\sigma \,d\sigma' \sum_{\alpha \dots \beta'}
 e^{-i(\sigma - \sigma')t}
\bigl( e^{-i{\rm H}_0 t} \bigr)^{\alpha \alpha' *}
 \bigl( e^{-i {\rm H}_0t }\bigr)^{\alpha \alpha''}\cdot \cr
&\qquad \cdot {_m^0\langle} 0 \alpha' \vert \sigma \beta \rangle_f^*
\,\,{ _m^0 \langle} 0 \alpha'' \vert \sigma' \beta'\rangle_f \,_f\langle
 \sigma \beta \vert \psi)^*
\,{_f\langle}\sigma' \beta' \vert \psi). \cr} \eqno(3.20)$$
Carrying out the sum over $\alpha$, the unitary factors cancel, leaving
 $\delta_{\alpha',\alpha''}$.
The $t$-integration then forms a factor $2\pi \delta (\sigma-\sigma')$,
 permitting a sum
on $\alpha' = \alpha''$. We show below that, from the unitarity of
 ${_f\langle}t \alpha \vert \sigma \beta \rangle_f$, it follows that the
indices in $_m\langle 0\alpha \vert
 \sigma \beta \rangle_f$
  label orthonormal sets in the auxiliary spaces attached
 to $t=0$ and $\sigma$,
 for each $\sigma$, i.e.,
$$ \sum_{\alpha}\,  _m^0\langle 0 \alpha' \vert \sigma \beta \rangle_f^*
\,\, { _m^0 \langle} 0 \alpha'
\vert \sigma \beta' \rangle_f = \delta_{\beta,\beta'},$$
and the therefore the final integral on $\sigma$ and sum on
 $\beta$ can be carried out in $(3.20)$:
$$\int d\sigma \sum_\beta \vert _f\langle
 \sigma \beta \vert \psi) \vert^2 = 1.$$
On the other hand, if $(3.19)$ were to provide the complete representation,
$$\sum_{\alpha, \alpha', \alpha''}
\bigl( e^{-i{\rm H}_0 t} \bigr)^{\alpha \alpha' *}
 \bigl( e^{-i {\rm H}_0t }\bigr)^{\alpha \alpha''}\,
{_m^0\langle} 0 \alpha' \vert \psi)^*\,\, { _m^0\langle} 0 \alpha'' \vert \psi)
= \sum_{\alpha'} \vert _m^0\langle 0 \alpha' \vert \psi) \vert^2 \eqno(3.21)$$
is bounded but independent of $t$; an integral over $t$ would then diverge.
\par We now remark that since
$$ _f\langle \sigma \beta \vert e^{-iK_0 \tau }\vert \psi) =
 e^{-i\sigma \tau}\,{ _f\langle} \sigma \beta
\vert \psi),  \eqno(3.22)$$
it follows from $(2.2)$ that
$$ \eqalign{{ _f\langle} t \beta \vert e^{-i K_0 \tau} \psi) &= \int d\sigma
 e^{i\sigma(t-\tau)}
\,{_f\langle} \sigma \beta \vert \psi) \cr
&= \,{_f\langle} t-\tau, \beta \vert \psi),\cr} \eqno(3.23)$$
making explicit the translation induced by $K_0$ in this representation, as
 is evident from $(2.1)$
(or the first of $(3.16)$).  It then follows that
$$  _f\langle t \beta \vert K_0 \vert \psi ) = -i\partial_t\,{ _f\langle} t
 \beta \vert \psi)
\eqno(3.24)$$
and $(3.16)$ becomes, in the free translation representation,
$$\eqalign{ {_m\langle} t \alpha \vert K_0 \vert t' \beta\rangle_f &=
 i\partial_{t'}\,
{_m\langle} t \alpha \vert t' \beta \rangle_f \cr
&= -i\partial_t \,{_m\langle} t \alpha \vert t' \beta\rangle_f +
 \sum_{\alpha'}
{\rm H}_0 ^{\alpha \alpha'}\,{ _m\langle} t \alpha' \vert t' \beta \rangle_f,
 \cr}  \eqno(3.25)$$
or
$$ (i\partial_t + \partial_{t'})\,{ _m\langle} t \alpha
\vert t' \beta \rangle_f =
\sum_{\alpha'} {\rm H}_0^{\alpha \alpha'}\,{ _m\langle} t
\alpha' \vert t' \beta
 \rangle_f.
\eqno(3.26)$$
It is clear from $(3.26)$ that the transformation function, ${_m\langle} t
\alpha
 \vert t' \beta\rangle_f$, from the representation in which
$T_m$ is diagonal,
$$ T_m = \int dt \sum_\alpha \vert t \alpha \rangle_m\, t\,\,\,{ _m\langle}
t\alpha
 \vert, \eqno(3.27)$$
to that for which the {\it free time operator}
$$ T_f = \int dt \sum_\beta \vert t\beta \rangle_f \,t \,\,\, {_f\langle} t
\beta
 \vert \eqno(3.28)$$
is diagonal, cannot be a function of $t-t'$ alone (in particular
, proportional to
$\delta(t-t')$), if the right hand side of $(3.26)$ is not zero.  We see
 that the existence
of a free Schr\"odinger type evolution operator, which can propagate a
 stable state, is necessary
for the non-trivially different structure of the free and model translation
 representations.
\par To find the general solution of $(3.26)$, let
$$ {_m\langle} t\alpha \vert t' \beta \rangle_f = f^{\alpha \beta} (t_+, t_-),
 \eqno(3.29)$$
where
$$ t_\pm = { t' \pm t \over 2}. \eqno(3.30)$$
Then, $(3.26)$ becomes
$$ i \partial_{t_+} f^{\alpha \beta } (t_+, t_-) = \sum_{\alpha'}
 {\rm H}_0 ^{\alpha
\alpha'}
f^{\alpha' \beta} (t_+, t_-)$$
with solution
$$ f^{\alpha \beta} (t_+, t_-) = \sum_{\alpha'} \bigl( e^{-i{\rm H}_0t_+}
 \bigr)^{\alpha\alpha'}
f^{\alpha' \beta}(0, t_-). \eqno(3.31)$$
It therefore follows that
$$\eqalign{ {_m\langle} t \alpha \vert \sigma \beta \rangle_f
 &= \sum_{\alpha'} \int dt' \,
e^{i\sigma t'} \bigl(e^{-i{\rm H}_0 t_+} \bigr)^{\alpha \alpha'}
f^{\alpha' \beta} (0, t_-) \cr
&= \sum_{\alpha' \alpha''} \int dt' e^{i\sigma t'} \bigl(e^{-i{\rm H}_0t}
 \bigr)^{\alpha
\alpha''} \bigl(e^{-i{\rm H}_0{t'-t \over 2}} \bigr)^{\alpha''
\alpha'}f^{\alpha' \beta }(0, t_-) \cr
&= \sum_{\alpha' \alpha''} \int d(t'-t) e^{i\sigma t }
 e^{i\sigma ( t'-t)} \bigl(e^{-i{\rm H}_0t} \bigr)^{\alpha
\alpha''} \bigl(e^{-i{\rm H}_0{t'-t \over 2}} \bigr)^{\alpha''
\alpha'}f^{\alpha' \beta }(0, t_-). \cr} \eqno(3.32)$$
We now define
$$ U^{\alpha \beta} (\sigma) = \sqrt{2\pi} \int dt\, e^{i\sigma t}
 \bigl(e^{-i{\rm H}_0t/2}
 \bigr)^{\alpha \alpha'}
f^{\alpha' \beta}(0, t/2) \eqno(3.33)$$
so that $(3.32)$ becomes
$$ _m\langle t \alpha \vert \sigma \beta \rangle_f = {1 \over \sqrt{2\pi}}
 \sum_{\alpha'}
e^{i\sigma t} \bigl( e^{-i{\rm H}_0 t}\bigr)^{\alpha \alpha'} U^{\alpha'
 \beta} (\sigma). \eqno(3.34)$$
It then follows that
$$ U^{\alpha \beta} (\sigma) = \sqrt{2\pi}\, {_m\langle}0 \alpha \vert
\sigma \beta \rangle_f . \eqno(3.35)$$
\par The unitarity relations for the transformation function
${_m\langle}t \alpha\vert\sigma \beta \rangle_f $
 imply the unitarity of $U^{\alpha \beta}(\sigma)$:
$$\eqalign{ \sum_\alpha \int dt\, {_f\langle}\sigma\beta\vert t
 \alpha \rangle_m \,{_m\langle}
t \alpha \vert \sigma' \beta' \rangle_f &= {1 \over {2\pi}}
 \sum_{\alpha \alpha'
\alpha''} \int dt\,e^{-i\sigma t} \bigl( e^{-i{\rm H}_0t}
 \bigr)^{\alpha\alpha'*}
U^{\alpha'\beta *}(\sigma) \cdot \cr
&\qquad \cdot e^{i\sigma' t} \bigl(e^{-i{\rm H}_0t} \bigr)^{\alpha \alpha''}
U_{\alpha''\beta'} \cr
&= \delta(\sigma - \sigma') \sum_\alpha U^{\alpha\beta *}
 (\sigma) U^{\alpha \beta'}
(\sigma) \cr} $$
so that
$$ \sum_\alpha U^{\alpha \beta *}(\sigma) U^{\alpha \beta'}(\sigma) =
\delta_{\beta\beta'}. \eqno(3.36)$$
Moreover,
$$ \eqalign{\sum_\beta &\int d\sigma\,{_m\langle} t \alpha \vert \sigma \beta
\rangle_f \,{_f\langle} \sigma \beta \vert t' \alpha' \rangle_m = \cr
&=  {1 \over {2\pi}}
\sum_{\beta \alpha'' \alpha'''} \int d\sigma\, e^{i\sigma(t-t')}
\bigl(e^{-i{\rm H}_0t} \bigr)^{\alpha \alpha''} \bigl(e^{-i{\rm H}_0 t'}\bigr)^
{\alpha' \alpha''' *}
\cdot U^{\alpha''\beta} (\sigma) U^{\alpha'''\beta *}(\sigma)\cr
&\qquad = \delta(t-t') \delta_{\alpha\alpha'}. \cr}  \eqno(3.37) $$

\par Now, suppose that ${\alpha, \alpha'}$ correspond to (generalized)
eigenstates
of ${\rm H}_0$; then, $(3.37)$ becomes
$$ \delta(t-t')\delta_{\alpha \alpha'} = {1 \over {2\pi}} \sum_\beta
\int d\sigma\, e^{i(\sigma -E_\alpha)t} e^{-i(\sigma - E_{\alpha'})t'}
U^{\alpha\beta}(\sigma)U^{\alpha'\beta *}(\sigma). \eqno(3.38)$$
Multiplying $(3.38)$ by $e^{-i\nu t}$ and integrating over $t$, we obtain
$$e^{-i\nu t'}\delta_{\alpha\alpha'} = e^{-i(\nu +E_\alpha - E_{\alpha'})t'}
\sum_\beta U^{\alpha\beta}(\sigma)U^{\alpha'\beta *}(\sigma)\vert_
{\sigma= E_\alpha + \nu} $$
for every $\nu$. This relation implies that $E_\alpha = E_{\alpha'}$,
 so  that
$$ \delta_{\alpha \alpha'} = \sum_\beta U^{\alpha\beta}(\sigma)
 U^{\alpha' \beta *}(\sigma).  \eqno(3.39)$$
\par The transformation function ${_m\langle} t \alpha \vert \sigma \beta
\rangle_f =
e^{i\sigma t}\, {_m^0\langle} t \alpha \vert \sigma \beta \rangle_f $
constitutes
a map from the spectral family associated with $T_m$, represented by the
 kets $\{\vert t\alpha \rangle_m \}$ to the spectral representation of $K_0$,
represented by the kets $\{\vert \sigma \beta \rangle_f \}$.  We can think of
this map
in two stages, the first from  $\{\vert t\alpha \rangle_m \}$ to a standard
frame
 $\{\vert \beta'\rangle_0 \}$ (projection) in the auxiliary space of the
 free representation, then a map (lift) from this to the foliated frames
  $\{\vert \sigma \beta \rangle_f \}$ according to
$$ {_m\langle} t \alpha \vert \sigma \beta \rangle_f = \sum_{\beta'}
{_m\langle} t \alpha \vert \beta'\rangle_0\,\,{_0\langle}\beta' \vert \sigma
\beta
\rangle_f  \eqno(3.40)$$
with the property $(3.17)$ due to the contraction with
 ${_0\langle}\beta' \vert \sigma \beta\rangle_f$.    Then, $(3.35)$ can be
written as
$$ U^{\alpha\beta}(\sigma) =  \sqrt{2\pi}\sum_{\beta'}{_m\langle}0\alpha \vert
\beta' \rangle_0 \,\, {_0\langle} \beta' \vert \sigma \beta \rangle_f .
 \eqno(3.41)$$
 Let us define the unitary map
$$ \langle \alpha \vert \beta' \rangle \equiv \sqrt{2\pi} {_m\langle}0\alpha
\vert \beta'\rangle_0, \eqno(3.42)$$
 so that
$$ \eqalign{U^{\beta'\beta} (\sigma) &\equiv {_0\langle} \beta' \vert \sigma
 \beta \rangle_f \cr
 &=\sum_{\alpha}
 \langle \beta' \vert \alpha \rangle
U^{\alpha \beta}(\sigma) \cr} \eqno(3.43)$$
corresponds to a transformation  in ``orientation'' of the representation
from the standard one, in the isomorphic auxiliary spaces.  The map $U^{\beta'
\beta}(\sigma)$ from a standard frame to a frame varying with $\sigma$ has the
geometric interpretation of a section of a frame bundle, as reflected in
$(3.40)$.
\bigskip
\noindent
{\bf 4. The $S$-Matrix for Pointwise Models.}
\smallskip
 \par In this section we define the Lax-Phillips wave operators for
the pointwise models discussed in the previous section,  and
compute the $S$-matrix (based on the intertwining of $K$ and $K_0$).  We
show that the Lax-Phillips $S$-matrix is, in this case, simply related to
 the $S$-matrix  of the usual scattering
problem (based on the intertwining of ${\rm H}$ and ${\rm H}_0$) by
the unitary operator
$U(\sigma)$.   This operator acts in a way similar  to that of the dilation
used by Aguilar and Combes$^{18}$ (see also Simon$^{19}$); analytic continuation
in $\sigma$ distorts the continuous spectrum of the Hamiltonian, exposing
the resonance poles on the first sheet.  We give an example, based on the
Lee-Friedrichs model$^{15}$ in the next section.
\par We  show in the following that the spectrally diagonal operator
$S^{\beta \beta'}(\sigma)$ for pointwise models has the form
$$ S^{\beta\beta'}(\sigma) = U^{\alpha \beta *}
 (\sigma)\bigl( S^{aux}\bigr)^{\alpha \alpha'}
U^{\alpha'\beta'}(\sigma). \eqno(4.1)$$
 Here,  $U^{\alpha \beta}(\sigma)$ is the operator on the auxiliary
space
 defined by $(3.35)$,
and $S^{aux}$ is the $S$-matrix of the usual scattering theory defined by
${\rm H},\,{\rm H}_0 $ in the auxiliary space.
\par To see this, we study the operator ${\bf S}$ in the form
$$ {\bf S} = \Omega_+^{-1}\Omega_- = \lim_{\tau \rightarrow \infty}
e^{iK_0\tau} e^{-2iK\tau} e^{iK_0\tau}, \eqno(4.2)$$
which can be expressed
as
$$\eqalign{{\bf S} &= \lim_{\epsilon\rightarrow 0} \epsilon \int_0^\infty
d\tau\, e^{-\epsilon \tau} e^{iK_0 \tau} e^{-2iK\tau} e^{iK_0\tau} \cr
&= \int_0^\infty d\tau \, \bigl(-{d \over d\tau}e^{-\epsilon\tau}\bigr)
e^{iK_0\tau} e^{-2iK\tau} e^{iK_0\tau} \cr
&= 1 - i\int_0^\infty d\tau\, \{ e^{iK_0\tau} V e^{-2iK\tau} e^{iK_0\tau} \cr
&\qquad + e^{iK_0\tau} e^{-2iK\tau} V e^{iK_0\tau} \} e^{-\epsilon\tau}.\cr}
\eqno(4.3)$$
\par In the free spectral representation, we therefore have
$$\eqalign{ {_f\langle}\sigma \beta \vert {\bf S}\vert \sigma' \beta' \rangle_f
&=
\delta(\sigma - \sigma') \delta^{\beta \beta'} \cr
&-i\int_0^\infty d\tau{_f\langle} \sigma \beta \vert V e^{i(\sigma + \sigma' -2K
+ i\epsilon)\tau} + e^{i(\sigma + \sigma' -2K
+ i\epsilon)\tau} V \vert \sigma' \beta' \rangle_f \cr
&= \delta(\sigma - \sigma') \delta^{\beta \beta'}\cr
& + {1 \over 2} {_f\langle} \sigma \beta
\vert V G({\sigma + \sigma' \over 2} + i\epsilon) + G({\sigma +
\sigma'
 \over 2} + i\epsilon)V
\vert \sigma' \beta' \rangle_f , \cr} \eqno(4.4)$$
where we use the definitions
$$ G(z) = { 1 \over z-K} , \qquad  G_0(z) = {1 \over z-K_0} . \eqno(4.5)$$
We now define the operator$^{20}$
$${\bf T}(z) = V + VG(z)V = V + VG_0(z) {\bf T}(z), \eqno(4.6)$$
where we have used the second resolvent equation
$$\eqalign{G(z) &= G_0(z) + G_0(z) V G(z) \cr
&= G_0(z) + G(z) V G_0(z).  \cr} \eqno(4.7)$$
Since
$$ \eqalign{ {\bf T}(z) G_0(z) &= V G_0(z) + VG(z)V G_0(z)  \cr
&= VG(z), \cr} \eqno(4.8)$$
and
$$ \eqalign{ G_0(z) {\bf T}(z) &= G_0(z) V + G_0(z)V G(z)V \cr
&= G(z) V, \cr } \eqno(4.9)$$
it follows that
$$\eqalign{{_f\langle} \sigma\beta \vert {\bf S} \vert \sigma' \beta' \rangle_f
&=
\delta(\sigma - \sigma') \delta^{\beta\beta'} \cr
& + {1 \over 2} {_f\langle} \sigma \beta \vert {\bf T}({\sigma + \sigma' \over
2}
 + i\epsilon)
G_0({\sigma + \sigma' \over 2} + i\epsilon) \cr
&\,\,\,\,\, + G_0({\sigma + \sigma' \over 2}
 + i\epsilon)
{\bf T}({\sigma + \sigma' \over 2} + i\epsilon) \vert \sigma' \beta'\rangle_f
\cr
&= \delta(\sigma - \sigma')\delta^{\beta \beta'}
 + \bigl\{ {1 \over \sigma - \sigma' +i\epsilon}
+ {1 \over \sigma' - \sigma + i\epsilon} \bigr\} {_f\langle} \sigma \beta \vert
{\bf T} ({\sigma + \sigma' \over 2} + i\epsilon) \vert \sigma' \beta' \rangle_f
\cr
&= \delta(\sigma- \sigma')\{ \delta^{\beta\beta'} - 2\pi i
\,{_f\langle}\sigma\beta
\vert{\bf T} (\sigma
+ i\epsilon) \vert \sigma \beta' \rangle_f \}. \cr } \eqno(4.10)$$
We remark that by this construction, we see that
$S^{\beta\beta'}(\sigma)$
 is {\it analytic in the upper half plane} in $\sigma$.
\par To complete our demonstration  of $(4.1)$, we expand ${\bf
T}(z)$ (aasuming that the series converges),
 using $(4.6)$, as
$${\bf T} = V + VG_0(z)V + VG_0(z)VG_0(z)V + \cdots \, . \eqno(4.11)$$
The matrix elements of ${\bf T}$ therefore involve
$$ {_f\langle} \sigma \beta \vert V \vert \sigma' \beta' \rangle_f =
\int dt\,
 \sum_{\alpha \alpha'} {_f\langle} \sigma \beta\vert t \alpha
\rangle_m
 V^{\alpha \alpha'} {_m\langle} t \alpha' \vert \sigma'
\beta'\rangle_f.
 \eqno(4.12)$$
  From $(3.34)$, we obtain
$$ {_f\langle}\sigma \beta \vert V \vert \sigma' \beta' \rangle_f =
 {1 \over 2\pi}\sum_{\alpha \alpha'} \int dt \, e^{i(\sigma' -
\sigma)t}
 U^{\alpha \beta *} (\sigma) V_I(t)^{\alpha \alpha'} U^{\alpha' \beta'
}
(\sigma'), \eqno(4.13) $$
where $V_I(t)$ is the interaction picture form for $V$ in the standard
 scattering theory,
$$V_I^{\alpha \alpha'} (t) = \sum_{\alpha'' \alpha'''}\bigl( e^{iH_0t}
 \bigr)^{\alpha \alpha''} V^{\alpha'' \alpha'''}(t) \bigl(e^{-iH_0t}
 \bigr)^{\alpha''' \alpha'}. \eqno(4.14)$$
It is convenient to write $(4.13)$ as an operator-valued kernel on
 the auxiliary space in the free spectral representation (suppressing
 the explicit indices of the auxiliary space), i.e.,
$$  {_f\langle} \sigma \vert V \vert \sigma' \rangle_f =
 {1 \over 2 \pi}\int dt \, e^{i(\sigma' - \sigma)t} U^\dagger
 (\sigma) V_I(t) U(\sigma'). \eqno(4.15)$$
Since
$$ {_f\langle} \sigma' \vert G_0(\sigma + i \epsilon) \vert \sigma''
 \rangle_f = {1 \over {\sigma - \sigma' + i \epsilon}} \delta(\sigma' -
\sigma''), $$
it follows that
$$\eqalign{ {_f\langle} \sigma \vert V G_0&(\sigma + i\epsilon) V
 \vert \sigma' \rangle_f = \int d\sigma''d\sigma''' \cdot \cr
&\,\,\,\,\cdot {_f\langle} \sigma \vert V \vert \sigma'' \rangle_f
 \,\, {_f \langle}
\sigma'' \vert G_0(\sigma + i\epsilon) \vert \sigma''' \rangle_f \,\,
 {_f\langle} \sigma''' \vert V \vert \sigma' \rangle \cr
&= U^\dagger(\sigma) {1 \over (2\pi)^2} \int d\sigma'' dt dt'
 {e^{i\sigma''(t-t')} \over {\sigma -\sigma'' + i\epsilon}}
 e^{-i\sigma t }e^{i\sigma' t'}  V_I(t) V_I(t') U(\sigma). \cr} \eqno(4.16)$$
Closing the contour in the upper half plane in $\sigma''$
 to include the pole at $\sigma'' = \sigma + i\epsilon$ requires
 $t>t'$ (for $t<t'$, the contour must be closed in the lower half
 plane and vanishes); the result,
for $t>t'$, is $-2\pi i e^{i(\sigma + i\epsilon)(t-t')}$, so that
$${_f\langle} \sigma \vert VG_0(\sigma + i\epsilon)V \vert \sigma'\rangle_f =
-{i \over 2\pi} U^\dagger(\sigma)\int_{-\infty}^{\infty} dt
\int_{-\infty}^t dt'
 e^{i(\sigma' - \sigma)t} V_I(t)V_I(t') U(\sigma'). \eqno(4.17)$$
  For $\sigma= \sigma'$, as enforced by $(4.10)$, the
exponential factor is unity.
\par  To see how the rest of the series goes, we calculate
$$ \eqalign{ {_f\langle} \sigma \vert V&G_0(\sigma+i\epsilon)
 V G_0(\sigma +i\epsilon)V \vert \sigma' \rangle_f = \cr
&= {1 \over (2\pi)^3}U^\dagger (\sigma) \int dt dt'dt'' d\sigma'' d\sigma'''
{e^{i(\sigma'' - \sigma)t}e^{i(\sigma''' - \sigma'')t'}
 e^{i(\sigma' - \sigma''')t''} \over (\sigma - \sigma''
 +i\epsilon)(\sigma - \sigma''' + i\epsilon)}\cdot \cr
&\,\,\,\,\,\,\,
\cdot V_I(t)V_I(t')V_I(t'')U(\sigma'), \cr} \eqno(4.18)$$
where the internal factors $U(\sigma''),\, U(\sigma''')$ cancel.
  Now, as above,
$$ \int d\sigma'' {e^{i\sigma''(t-t')} \over {\sigma - \sigma''
 +i\epsilon}} = -2 \pi i e^{i\sigma(t-t')} \,\,\,\, t>t', $$
and is otherwise zero.  The integral over $\sigma'''$ then yields
$$ \int d\sigma''' {e^{i\sigma'''(t'-t'')} \over {\sigma -  \sigma'''
 + i\epsilon}} = -2 \pi i e^{i\sigma(t'-t'')} \,\,\,\, t'>t'',$$
and is otherwise zero, so we conclude that a non-zero result requires
 $t>t'>t''$, and in this case
$$ \eqalign{{_f\langle} \sigma \vert V &G_0(\sigma
+i\epsilon)VG_0(\sigma +
 i\epsilon)V \vert \sigma' \rangle_f = \cr
&= {i^2 \over 2\pi} U^\dagger (\sigma)\int_{-\infty}^\infty \,
dt\int_{\infty}^t \, dt'
\int_{-\infty}^{t'}\,  dt'' \,
 V_I(t) V_I(t')V_I(t'') U(\sigma') e^{i(\sigma' - \sigma)t''} ; \cr}
 \eqno(4.19)$$
the last factor again becomes unity under the restriction $\sigma =
\sigma'$.   The general result for the series is
$$\eqalign{ {_f\langle} \sigma \vert {\bf S} \vert \sigma' \rangle_f &=
\delta(\sigma - \sigma') U^\dagger(\sigma)\bigl\{ 1 -
 i\int_{-\infty}^\infty dt\, V_I(t) \cr
&+{(-i)^2 \over 2!} {\cal T}\int_{-\infty}^\infty dt dt' \,
 V_I(t)V_I(t') \cr
&+{ (-i)^3\over 3!} {\cal T} \int_{-\infty}^\infty dt dt'dt'' \,
 V_I(t)V_I(t')V_I(t''') \cr &+ \cdots \bigr\} U(\sigma), \cr}
\eqno(4.20)$$
where ${\cal T}$ indicates that the operations must be time-ordered
under the integrals.
The terms in the bracket in $(4.20)$ are the expansion of
$$ S^{aux} = {\cal T} \biggl( e^{-i\int_{-\infty}^\infty V_I(t)dt }
 \biggr), \eqno(4.21)$$
 so that $(4.1)$ is proven.
\par We have constructed the incoming and outgoing subspaces
 ${\cal D}_\pm$ in $(2.6)$.  It is essential for application
 of the Lax-Phillips theory that these subspaces be orthogonal,
 i.e., for every $f_+ \in {\cal D}_+, \, f_- \in {\cal D}_-,$ that
 $(f_+, f_-) = 0$.  If
$$ \eqalign{ f_+ &= \Omega_+ f_0^+ \cr
f_- &= \Omega_- f_0^- , \cr} \eqno(4.22)$$
mapped from functions in ${\cal D}_0^\pm$, we see that the orthogonality
condition is
$$ (f_+, f_-) = (f_0^+, \Omega_+^{-1} \Omega_- f_0^-) = 0. \eqno(4.23)$$
We now show that the $S$-matrix leaves the support of the
 functions in ${\cal D}_-$ in the incoming representation invariant,$^1$ and
therefore  the orthogonality condition is satisfied.  As shown in $(2.11)$,
 the $S$-matrix in free representation transforms the incoming to
 the outgoing representation; we may therefore write the scalar
product in
 $(4.23)$ as
$$(f_+, f_-) = \sum_{\beta \beta'} \int dt dt' \, (f_0^+ \vert t
 \beta \rangle_{out} \,\,{_f\langle} t \beta \vert {\bf S} \vert t'
 \beta' \rangle_f \,\, {_{in}\langle} t' \beta' \vert f_0^- )  \eqno(4.24)$$
Now,
$$\eqalign{{_f\langle} t \beta \vert {\bf S} \vert t' \beta' \rangle_f &=
     \int d\sigma d\sigma' \, e^{i\sigma t} e^{-i\sigma' t'}
     {_f\langle}\sigma \beta \vert {\bf S} \vert \sigma' \beta' \rangle_f \cr
&= \int d\sigma e^{i\sigma(t-t')}  S^{\beta \beta'}(\sigma) \cr
&= S^{\beta \beta'} (t-t').  \cr } \eqno(4.25)$$
The  function $S(\sigma)^{\beta \beta'}$ is analytic in the upper
 half plane; it may have a null co-space, but is otherwise regular.
  Its singularity lies in the lower half plane .
  To find the non-vanishing value for  $S^{\beta \beta'}(t-t')$,
 we must close the contour in the lower half plane.
 This can only be done if $t' >t$.  For $t' <t$,
 one must close in the upper half plane, and there
 $S(\sigma)$ has no singularity, so the integral vanishes.  Hence
$S^{\beta \beta'}(t-t')$ takes ${\cal D}_-$ to ${\cal D}_-$
 in the incoming representation, and the subspaces ${\cal D}_+$
 and ${\cal D}_-$ are orthogonal.
\par We finally remark that the $S$-matrix, in the {\it model} space,
 has the form
$$ \eqalign{ {_m\langle} t \alpha \vert {\bf S} \vert t'
 \alpha'\rangle_m &= \sum_{\beta \beta'} \int d\sigma d\sigma' \,
 {_m\langle} t \alpha \vert \sigma \beta \rangle_f \,\, {_f\langle}
 \sigma \beta \vert {\bf S} \vert \sigma' \beta' \rangle_f \,\,
 {_f\langle} \sigma' \beta' \vert t' \alpha' \rangle_m \cr
&= \sum _{\beta \beta'} \int d\sigma \, {_m\langle}t \alpha \vert
 \sigma \beta \rangle_f \, U^{\dagger \beta \alpha}(\sigma)
 S^{aux \, ,  \alpha \alpha'} U^{\alpha' \beta'} (\sigma)
 {_f\langle} \sigma \beta' \vert t' \alpha' \rangle_m \cr
&= {1 \over 2 \pi} \int d\sigma e^{i\sigma(t-t')}
 \bigl( e^{-iH_0 t} \bigr) ^{\alpha \alpha''} U^{\alpha''
 \beta}(\sigma) U^{\dagger \beta \alpha'''}(\sigma)
 \cdot \cr &\,\,\,\,\cdot S^{aux \, ,  \alpha'''
  \alpha^{iv}} U^{\alpha^{iv}\beta'}(\sigma)
 U^{\dagger \beta' \alpha^{v}}(\sigma)
 \bigl(e^{-iH_0 t'}\bigr)^{ \alpha^{v}\alpha' *} \cr
&= \delta(t-t') S^{{aux}\, , \alpha \alpha'} , \cr} \eqno(4.26)$$
where we have used $(3.34)$ and the fact that $H_0$ commutes with $S^{aux}$.
In the model space, $S^{aux}$ acts at a
 given $t$,
 and multiplication by $\delta(t-t')$ constitutes
 the lift of this operator to the Lax-Phillips theory. This result
 illustrates the conclusion of ref. 4, that for a Hamiltonian that is
 pointwise in $t$,
 the Lax-Phillips $S$-matrix has no non-trivial analytic structure in
 the model representation.  In the free spectral representation, however,
 it has the non-trivial analytic structure necessary for establishing the
 relation between the singularities of $S(\sigma)$ and the spectrum of
 the generator of the semigroup.
\bigskip
\noindent
{\bf 5. The Lee-Friedrichs Model}
\smallskip
\par In this section, we work out a specific illustrative
 example, the well-known time independent soluble model of
 Friedrichs and Lee$^{15}$.
We show that, for a simple choice of $U(\sigma)$,
 the exponential Lax-Phillips decay law can coincide
 with the Lee-Friedrichs pole approximation.
\par The Lee-Friedrichs model for scattering and resonances$^{15,21}$,
 in the framework of standard non-relativistic scattering theory, is
characterized by
a Hamiltonian  ${\rm H}= {\rm H}_0 + V$ for which ${\rm H}_0$
 has a bound state with eigenfunction $\phi$ and eigenvalue $E_0$
 embedded in an absolutely continuous spectrum on $(0, \infty)$,
 and for which $V$ has matrix elements only from the discrete
 bound state to the generalized eigenfunctions on the continuum.
The vanishing of continuum-continuum matrix elements corresponds
 to the assumption, often a good approximation, that there are
 no final state interactions.  For the time independent case,
 which we treat here, the Lax-Phillips $S$-matrix, according
 to $(4.1)$, can be written as
$$ S(\sigma) = \lim_{t \rightarrow \infty} e^{i{\rm H}_0(\sigma)t}
 e^{-2i{\rm H}(\sigma)t}e^{i{\rm H}_0(\sigma)t}, \eqno(5.1)$$
where ${\rm H}(\sigma) = U^\dagger(\sigma) {\rm H} U(\sigma)$
 and $ {\rm H}_0(\sigma) = U^\dagger(\sigma) {\rm H}_0 U(\sigma)$.
 We then construct $U(\sigma)$ so that it induces a
 diffeomorphism\footnote{*}{We thank G. Goldin for a
 discussion of this point.} on the spectrum of the
 unperturbed Hamiltonian in such a way that, for ${\rm H}_0 \vert
 \lambda \rangle = \lambda \vert \lambda \rangle$,
$$ U(\sigma) \vert \lambda \rangle = \vert \lambda \rangle_\sigma
 = \vert \lambda(\sigma)\rangle \sqrt{\Lambda(\sigma, \lambda)}, \eqno(5.2)$$
where
$$ \Lambda(\sigma, \lambda) = {d\lambda(\sigma) \over d\sigma} . \eqno(5.3)$$
The factor of the square root of the Jacobian is necessary
 to assure that $\langle \lambda \vert \lambda' \rangle
  = \delta (\lambda - \lambda') \equiv \langle \lambda
 \vert U^\dagger (\sigma) U(\sigma) \vert \lambda'\rangle$.
 It follows that  ${\rm H}_0(\sigma) \vert \lambda \rangle =
 \lambda(\sigma)\vert \lambda \rangle$, where $\lambda
 \rightarrow \lambda(\sigma)$ is a smooth map which can
 be analytically continued.  The procedure outlined above
 in $(4.2)$-$(4.10)$, applied to the auxiliary space problem,
 then results in
$$\eqalign{\langle \lambda \vert &S(\sigma) \vert \lambda'
 \rangle = \delta(\lambda - \lambda') \cr
& -2\pi i \delta (\lambda(\sigma) - \lambda'(\sigma) )
 \langle \lambda \vert T_\sigma^{aux} (\lambda(\sigma) +
 i\epsilon) \vert \lambda' \rangle, \cr}
\eqno(5.4)$$
where
$$ \eqalign{T_\sigma^{aux} (z) &= V_\sigma +
 V_\sigma G_\sigma^{aux} (z) V_\sigma,\cr
V_\sigma &= U^\dagger(\sigma) V U(\sigma) \cr} \eqno(5.5)$$
and
$$ G_\sigma^{aux} (z) = U^\dagger(\sigma) G^{aux}(z) U(\sigma). \eqno(5.6)$$
Here, $G^{aux}(z) = (z-H)^{-1}$ is a resolvent kernel
 in the auxiliary space, and we have written $(5.4)$
 in terms of operator-valued matrix elements, suppressing
 the degeneracy indices (such as angular variables).
\par Since $\lambda \rightarrow \lambda(\sigma)$ is $1:1$
 for each $\sigma$, we may write $(5.4)$ as
$$ \langle \lambda \vert S(\sigma)\vert \lambda' \rangle =
 \delta(\lambda - \lambda') \bigl\{ 1 - {2 \pi i \over
 \Lambda(\sigma, \lambda)} \langle \lambda \vert T_\sigma
 (\lambda(\sigma) + i\epsilon) \vert \lambda \rangle \bigr\}, \eqno(5.7) $$
The relations $(5.1)$-$(5.7)$ are valid for any
 $t$-independent non-relativistic scattering problem (for which $U(\sigma)$
induces
a diffeomorphism).  For the Lee-Friedrichs$^{15}$ model,
 in particular, since the set $\{ \vert \lambda(\sigma)
 \rangle \}$ is complete on the continuum, if we assume
 that the bound state embedded in the continuum is
 non-degenerate, the action of $U(\sigma)$ in the
 discrete eigenstate is just a phase, i.e.,
$$ U(\sigma) \phi = u_0(\sigma) \phi, \eqno(5.8)$$
where $u_0(\sigma)$ is a smooth complex valued function
 of $\sigma$ with unity modulus.  Then, since in this model,
 $\langle \lambda \vert V_\sigma \vert \lambda'\rangle = 0 $,
$$ \langle \lambda \vert V_\sigma G_\sigma (\lambda(\sigma) +
 i\epsilon) V_\sigma \vert \lambda \rangle =
 \Lambda(\sigma,\lambda) W(\lambda (\sigma))
 R_\sigma (\lambda (\sigma) + i\epsilon), \eqno(5.9)$$
where (the modulus squared implies a sum over degeneracy
 parameters as well)\footnote{\S}{We use round brackets
 for scalar products with proper (auxiliary) Hilbert
 space vectors, and angular brackets for generalized
 states on the continuum.}
$$ W(\lambda(\sigma)) = (\Lambda(\sigma, \lambda))^{-1}\vert
 \langle \lambda \vert V_\sigma \vert \phi)\vert^2 =
 \vert \langle \lambda(\sigma)\vert V \vert \phi)\vert^2 \eqno(5.10)$$
and
$$ R_\sigma(z) = \bigl( \phi \vert {1 \over \lambda(\sigma)
 - H(\sigma) + i\epsilon} \vert \phi \bigr) \equiv
 (\phi \vert G_\sigma(\lambda(\sigma) + i\epsilon) \vert \phi). \eqno(5.11)$$
 Using the second resolvent equation in the form
$$ G_\sigma(z) = G_\sigma^0(z) + G_\sigma^0(z) V_\sigma G_\sigma(z),
\eqno(5.12)$$
where
$$ G_\sigma^0 (z) =
 {1 \over \lambda(\sigma) - H_0(\sigma) + i\epsilon} , \eqno(5.13)$$
one obtains
$$\eqalign{ (\phi \vert &G_\sigma(\lambda(\sigma) +
 i\epsilon) \vert \phi) = \cr &= {1 \over \lambda(\sigma) -E_0 +i\epsilon} +
{1 \over \lambda(\sigma) -E_0 +i\epsilon}\int d\lambda' \,
 (\phi \vert V_\sigma \vert \lambda' \rangle \, \langle
 \lambda' \vert G_\sigma (\lambda(\sigma) + i \epsilon)
 \vert \phi) \cr } \eqno(5.14)$$
and
$$ \langle \lambda' \vert G_\sigma (\lambda(\sigma) +
 i\epsilon) \vert \phi) + {1 \over \lambda(\sigma)
 -\lambda'(\sigma) + i\epsilon} \langle \lambda'
 \vert V_\sigma\vert \phi) (\phi \vert G_\sigma(\lambda(\sigma)
 + i \epsilon) \vert \phi). \eqno(5.15)$$
Substituting $(5.15)$ into $(5.14)$, we obtain a formula for the reduced
resolvent
$$ \eqalign{R_\sigma (\lambda(\sigma) + i \epsilon) &=
 \bigl( \lambda(\sigma) +i\epsilon - E_0 - \int d\lambda' \,
 \Lambda(\sigma, \lambda) {W(\lambda'(\sigma)) \over
 \lambda(\sigma) - \lambda'(\sigma) + i\epsilon} \bigr)^{-1} \cr
&= \bigl(\lambda(\sigma) +i\epsilon - E_0 - \int d\lambda'(\sigma) \,
 {W(\lambda'(\sigma)) \over \lambda(\sigma) -
 \lambda'(\sigma) + i\epsilon} \bigr)^{-1}, \cr } \eqno(5.16)$$
since
 $$d\lambda' \Lambda(\sigma, \lambda') =  d\lambda'(\sigma)$$
The condition  $S(\sigma)n(\lambda)$ have a co-dimension in the
auxiliary space
is then that there
 exist some (measurable) $m(\lambda)$,
 such that (the factor
 $\Lambda^{-1}$ in $(5.7)$ cancels the factor $\Lambda$ of $(5.9)$)
$$ \int\,d\lambda d\lambda' \, m(\lambda)^*\langle \lambda \vert
S(\sigma) \vert \lambda' \rangle
n(\lambda') = 0,  \eqno(5.17)$$
  for all $n(\lambda)$.  Hence,
$$ \bigl\{ 1 - 2 \pi i \, W(\lambda(\sigma))
 R_\sigma(\lambda(\sigma) + i \epsilon)) \bigr\}^* m(\lambda) = 0.
\eqno(5.18)$$
For the case of a non-degenerate spectral model, for which
 $m(\lambda)$ is a one-dimensional function of $\lambda$,
 for every $\lambda$ in the support of $m(\lambda)$, the expression in
brackets must vanish, i.e.,
we must have
$$ \lambda(\sigma) +i\epsilon - E_0 - \int_0^\infty  d\lambda'(\sigma)
 \, {W(\lambda'(\sigma)) \over \lambda(\sigma) - \lambda'(\sigma)
 + i\epsilon} = 2 \pi i \, W(\lambda(\sigma)). \eqno(5.19)$$
Changing variables, $\lambda'(\sigma) \rightarrow \lambda'$,
 we can write the condition $(5.19)$ as
$$ \lambda(\sigma) +i\epsilon- E_0 - \int_0^\infty d\lambda' \,
 {W(\lambda') \over \lambda(\sigma) - \lambda' +i\epsilon
} = 2 \pi i W(\lambda(\sigma)). \eqno(5.20)$$
Note that the right hand side would precisely cancel with the
 jump contribution if we crossed the cut in the integral from
 above, interpreting $\lambda(\sigma) \rightarrow \zeta$
 as a complex variable. Taking the imaginary part of $(5.20)$ in the form
$$ \zeta +i\epsilon- E_0 - \int_0^\infty d\lambda' \,
 {W(\lambda') \over (\zeta - \lambda')} = 2 \pi i W(\zeta), \eqno(5.21)$$
one finds, as the the usual treatment of the Lee-Friedrichs
 model$^{21}$ (numerical studies were carried out in ref. 22),
$$ {\rm Im }
\zeta \bigl( 1 + \int d\lambda' \,{W(\lambda')
 \over |\zeta - \lambda'|^2} \bigr) \cong W(\zeta), \eqno(5.22)$$
where we have assumed ${\rm Im}\zeta$ small, and
 $W(\zeta)$ analytic in some neighborhood of the
 real axis, and primarily real for ${\rm Im}\zeta$ small.
The Eq. $(5.22)$ therefore may have a  solution
 in the first sheet of the upper half plane.
 As in dilation analytic theories$^{18,19}$,
 the Lax-Phillips method has effectively moved the cut below the root.
  For the value of $\sigma$ admitting a null space,
 and for a value of $\lambda = \lambda_0$ that
 satisfies $(5.20)$, there must be a finite
 interval in the neighborhood of $\lambda_0$
 that also satisfies this condition so that
 $m(\lambda)$ can have measurable support.
 Hence, the first derivative of the condition $(5.20)$ must be valid.
\par Taking the derivative of $(5.20)$, one obtains
$$ \Lambda(\sigma, \lambda)\bigl\{1 + \int d\lambda'
 {W(\lambda') \over (\lambda(\sigma) - \lambda')^2}
 - 2\pi i W'(\lambda(\sigma)) \bigr\} = 0. \eqno(5.23)$$
For sufficiently small coupling (in the generic case,
 for which the expression in brackets does not,
 by a special analytic structure, vanish), reflected
 by the size of $W(\lambda)$, the condition $(5.23)$
 cannot be satisfied unless the map induced by
 $U(\sigma)$ has a critical point, i.e.,
$$\Lambda(\sigma, \lambda) = {d\lambda(\sigma) \over d\lambda }=0 \eqno(5.24)$$
at the point $\zeta$ satisfying the condition $(5.21)$.
\par The function $\lambda(\sigma)$ is a monotonic
 and smooth function on the real line, but its
 analytic continuation in $\sigma$ can have such properties.
For $\lambda(\sigma)$ in the neighborhood of the solution
 $\zeta_0$ of $(5.21)$, the form
$$\lambda(\sigma) \cong {\lambda_0^2 \over \sigma}
 + {\rm O}((\lambda - \lambda_0)^2), \eqno(5.25)$$
satisfies the conditions $(5.21)$ and $(5.24)$ for
 $\sigma = \zeta_0^*$, $\lambda = \lambda_0 = |\zeta_0|$.  In this
simple one-dimensional illustration, the function $m(\lambda)$ is
 characterized only by its local support property on the
 spectrum of ${\rm H}_0$.  It would then follow from
 $(5.20)$ and $(5.22)$ that the Lax-Phillips
 exponential law goes precisely with the pole of the Lee-Friedrichs model.
\bigskip
\noindent
{\bf 6. Conclusions and Discussion}
\smallskip
\par An exact semigroup evolution law (exponential decay), corresponding
to an irreversible process, can be achieved within the
 framework of a microscopic quantum theory if the Hilbert
 space carries a natural foliation along an axis in its
 measure space on which the wave function moves by
 translation, under the full unitary evolution, in
 a special class of (translation) representations.  The
 foliation of such a space is assured by a theorem of Sinai$^{14}$
 when there are distinguished incoming and outgoing
 subspaces ${\cal D}_\pm$ which are stable under forward
 (backward) unitary evolution.  Lax and Phillips developed
 a complete theory of such systems
for the case of classical hyperbolic (wave) equations for
 scattering on a bounded target$^1$.  Flesia and Piron$^2$
 showed that the quantum mechanical Hilbert space can be
 extended, by a direct integral construction over the
 time variable, to form a structure in which the
 Lax-Phillips theory can be applied.  In a succeeding study$^4$,
 it was shown that a necessary condition for a non-trivial
 Lax-Phillips theory, for which the singularities of
 the $S$-matrix in the spectral variable constitute the spectrum
 of the generator of the semigroup, is that the evolution operator
 act as a smooth (operator-valued) integral kernel on the time axis
 in the free translation representation. We have shown in this paper
 that a  {\it pointwise} (in $t$) dynamical evolution operator in
 what we have called the model representation, in which the
 Hamiltonian of a system and the time variable appear with
 their usual laboratory interpretation, maps into a smooth,
 non-trivial kernel in the free translation representation, and
 therefore satisfies this necessary condition.
\par  We have shown that the subspaces ${\cal D}_\pm$ may be
 constructed from the wave operators, intertwining the full
 and unperturbed Lax-Phillips evolution operators, applied
 to functions with definite half-line support properties
 on the $t$-axis.  The orthogonality of these subspaces follows
 from the analytic properties of the $S$-matrix.
\par We have furthermore shown that the Lax-Phillips $S$-matrix
 is equivalent to the $S$-matrix of the standard
 scattering theory (for the time dependent case as well) by
 a unitary transformation  which is parametrized by the
 Lax-Phillips spectral variable.  This unitary transformation
 arises from the transformation from the model representation
 to the free spectral representation (the Fourier transform of
 the free translation representation).
There is considerable freedom in choosing such a function,
 which has the property, upon analytic continuation to the upper
 half-plane, of bringing the $S$-matrix to a form in which there
 is a non-trivial null co-space, corresponding to the eigenvectors
 of the resonant state (these points are conjugate to the resonant
 poles in the lower half plane). Since these vectors lie in the
 (auxiliary) Hilbert space, they may be used to compute expectation
 values of the usual dynamical variables, such as position, momentum,
 or angular momentum. Such properties are not available for
 the generalized functions obtained in the method of constructing
 Gel'fand triples$^{13}$ or the dilation analytic methods$^{18,19}$.
\par  The work of Lee, Oehme and Yang$^9$ and Wu and Yang$^9$,
 assuming an effective Hamiltonian analogous to the Wigner-Weisskopf pole
 approximation in the form of a two-by-two non-Hermitian matrix,
 results in an exact semigroup structure.
 As has been pointed out$^8$, deviations due to a more accurate
 treatment of the Wigner-Weisskopf method, reflecting its
 non-semigroup structure, could be important in regeneration
 processes; if, however, as the experimental results on $K$-meson
 decay$^{11}$ seem to imply, the phenomenological parameterization
 of refs. 9 are indeed consistent to a high level of accuracy,
 the Lax-Phillips theory could provide a microscopic framework
 for this exact semigroup behavior.
\par We gave here an illustration of the method for a one channel
 non-degenerate Lee-Friedrichs model$^{15,21}$ for the underlying dynamics.
The illustration was worked out by assuming, for simplicity,
  that the unitary transformation relating the Lax-Phillips
 $S$-matrix and that of the usual quantum scattering induces
 a diffeomorphism on the continuous spectrum without altering
 the spectral family; upon analytic continuation, we showed that
 the resulting Lax-Phillips semigroup can be generated by
 the complex pole of the usual Lee-Friedrichs model, and a
 measurable eigenfunction in the Hilbert space can be found
 for the null co-
space of the $S$-matrix at this point.  A more
 detailed analysis of this model, as well as other applications,
 for example,  to the two channel problem (e.g., $K$ meson decay),
 atomic and molecular and condensed matter physics, will be discussed
elsewhere.
\par We finally remark that the $\tau$-dependent decay law of the
unstable system can be calculated, as we have pointed out in
Eq.$(1.9)$, in terms of the full evolution acting on a state in ${\cal
K}$. The
Laplace transform of this amplitude then exhibits the full Green's
function $(z-K)^{-1}$, which can be expanded in an infinite series,
 with the help of $(4.7)$, as was done for the $S$-matrix in Section
4. The Laplace transform of the amplitude can then be expressed in
terms of the Lax-Phillips $S$-matrix, realizing the theorem of Lax and
Phillips$^1$ cited in Section 1.  The details of the demonstration of
this result and its applications will also be given elsewhere.

\bigskip
\noindent
\smallskip
{\it Acknowledgements}
\par One of us (L.H.) wishes to thank S.L. Adler, C. Newton,
 C. Piron and T.T. Wu for the opportunity to discuss the structure
 of the Lax-Phillips theory (and to express his appreciation to
 Professor Adler for his
 hospitality at the Institute for Advanced Study where the work
 reported here was started), B. Altshuler and O. Agam for their
 encouragement and discussions of possible applications to condensed
 matter physics, and I. Dunietz , Y.B. Hsiung, and B. Winstein
 and other colleagues at the Fermilab for many discussions of the
 implications of the current experiments on $K$ meson decay as well
 as potential applications in $B$ physics
for the theoretical structure that we have developed here.

\bigskip
\noindent
\frenchspacing
{\bf References}
\item{1.} P.D. Lax and R.S. Phillips, {\it Scattering Theory},
Academic Press,
 N.Y. (1967).
\item{2.} C. Flesia and C. Piron, Helv. Phys. Acta {\bf 57}, 697
(1984).
\item{3.} L.P. Horwitz and C. Piron, Helv. Phys. Acta {\bf 66}, 694 (1993).
\item{4.} E. Eisenberg and L.P. Horwitz, in {\it Advances in
 Chemical Physics}, {\bf XCIX}, p. 245, ed. I. Prigogine and
 S. Rice, John Wiley and Sons, N.Y. (1997).
\item{5.} C. Piron, {\it Foundations of Quantum Physics},
 Benjamin/Cummings, Reading, Mass. (1976).
\item{6.} V.F. Weisskopf and E.P. Wigner, Zeits. f. Phys.
 {\bf 63}, 54 (1930); {\bf 65}, 18 (1930).
\item{7.} L.P. Horwitz, J.P. Marchand and J. LaVita,
 J. Math. Phys. {\bf 12}, 2537 (1971); D. Williams,
 Comm. Math. Phys. {\bf 21}, 314 (1971).
\item{8.} L.P. Horwitz and L. Mizrachi, Nuovo Cimento {\bf 21A}, 625 (1974).
\item{9.} T.D. Lee, R. Oehme and C.N. Yang, Phys. Rev.
 {\bf 106}, 340 (1957); T.T. Wu and C.N. Yang, Phys. Rev. Lett. {\bf
13}, 380
 (1964).
\item{10.}G. Gamow, Z. Phys. {\bf 51}, 204(1928).
\item{11.} B. Winstein, {\it et al}, {\it Results from the
 Neutral Kaon Program at Fermilab's Meson Center Beamline, 1985-1997},
 FERMILAB-Pub-97/087-E, published on behalf of the E731, E773
 and E799 Collaborations, Fermi National Accelerator Laboratory,
 P.O. Box 500, Batavia, Illinois 60510.
\item{12.} S.R Wilkinson, C.F. Bharucha, M.C. Fischer, K.W. Madison,
P.R. Morrow, Q. Niu, B. Sundaram, and M. Raizen, Nature {\bf 387}, 575
(1997)
\item{13.}  W. Baumgartel, Math. Nachr. {\bf 69}, 107 (1975);
 L.P. Horwitz and I.M. Sigal, Helv. Phys. Acta
 {\bf 51}, 685 (1978); G. Parravicini, V. Gorini
 and E.C.G. Sudarshan, J. Math. Phys. {\bf 21}, 2208
 (1980); A. Bohm, {\it Quantum Mechanics: Foundations
 and Applications\/,} Springer, Berlin (1986); A. Bohm,  M. Gadella
and
 G.B. Mainland, Am. J. Phys. {\bf 57}, 1105 (1989); T. Bailey and
 W.C. Schieve, Nuovo Cimento {\bf 47A}, 231 (1978).
\item{14.} I.P. Cornfield, S.V. Formin and Ya. G. Sinai,
{\it Ergodic Theory}, Springer, Berlin (1982).
\item{15.} K.O. Friedrichs, Comm. Pure Appl. Math. {\bf 1},
 361 (1948); T.D. Lee, Phys. Rev. {\bf 95}, 1329 (1956).
\item{16.} J.S. Howland, {\it Springer Lecture Notes in Physics,},
 vol. 130, p. 163, Springer, Berlin (1979).
\item{17.} T. Kato, {\it Perturbation Theory for Linear Operators},
 Springer, N.Y. (1966).
\item{18.} J. Aguilar and J.M. Combes, Comm. Math. Phys. {\bf 22},
 280 (1971); J.M. Combes, Proc. Int. Cong. of Math., Vancouver (1974).
\item{19.} B. Simon, Ann. Math. {\bf 97}, 247 (1973).
\item{20.} For example, J.R. Taylor, {\it Scattering Theory},
 John Wiley and Sons,
 N.Y. (1972); R.J. Newton, {\it Scattering Theory of Particles
 and Waves}, McGraw Hill, N.Y. (1976).
\item{21.} L.P. Horwitz and J.P. Marchand, Rocky Mountain J. of Math.
 {\bf 1}, 225 (1971).
\item{22.} N. Bleistein, H. Neumann, R. Handelsman and L.P. Horwitz,
 Nuovo Cimento {\bf 41A}, 389 (1977).

\vfill
\end
\bye